\documentclass{wscpaperproc}
\usepackage{latexsym}
\usepackage{graphicx}
\usepackage{mathptmx}
\usepackage[T1]{fontenc}
\usepackage{pdflscape}
\usepackage{tabularx}

\usepackage{amsmath}
\usepackage{amsfonts}
\usepackage{amssymb}
\usepackage{amsbsy}
\usepackage{amsthm}

\usepackage[pdftex,colorlinks=true,urlcolor=blue,citecolor=black,anchorcolor=black,linkcolor=black]{hyperref}

\newtheoremstyle{wsc}
{3pt}
{3pt}
{}
{}
{\bf}
{}
{.5em}
{}

\theoremstyle{wsc}

\begin{document}

%
%

\pagestyle{fancyplain}

\thispagestyle{plain}
\firstPageHead{}

\chead{\fancyplain{}{\itshape Farr, Ng, Prochaska, Cruickshank, and West}}

\rhead{}
\cfoot{}
\renewcommand{\headrulewidth}{0pt} 

\makeatletter
\let\@internalcite\cite
\def\cite{\def\@citeseppen{-1000}%
    \def\@cite##1##2{(##1\if@tempswa , ##2\fi)}%
    \def\citeauthoryear##1##2##3{##1 ##3}\@internalcite}
\def\citeNP{\def\@citeseppen{-1000}%
    \def\@cite##1##2{##1\if@tempswa , ##2\fi}%
    \def\citeauthoryear##1##2##3{##1 ##3}\@internalcite}
\def\citeN{\def\@citeseppen{-1000}%
    \def\@cite##1##2{##1\if@tempswa, ##2)\else{}\fi}%
    \def\citeauthoryear##1##2##3{##1 (##3)}\@citedata}
\def\citeA{\def\@citeseppen{-1000}%
    \def\@cite##1##2{(##1\if@tempswa , ##2\fi)}%
    \def\citeauthoryear##1##2##3{##1}\@internalcite}
\def\citeANP{\def\@citeseppen{-1000}%
    \def\@cite##1##2{##1\if@tempswa , ##2\fi}%
    \def\citeauthoryear##1##2##3{##1}\@internalcite}
\def\shortcite{\def\@citeseppen{-1000}%
    \def\@cite##1##2{(##1\if@tempswa , ##2\fi)}%
    \def\citeauthoryear##1##2##3{##2 ##3}\@internalcite}
\def\shortciteNP{\def\@citeseppen{-1000}%
    \def\@cite##1##2{##1\if@tempswa , ##2\fi}%
    \def\citeauthoryear##1##2##3{##2 ##3}\@internalcite}
\def\shortciteN{\def\@citeseppen{-1000}%
    \def\@cite##1##2{##1\if@tempswa, ##2\else{}\fi}%
    \def\citeauthoryear##1##2##3{##2 (##3)}\@citedata}
\def\shortciteA{\def\@citeseppen{-1000}%
    \def\@cite##1##2{(##1\if@tempswa , ##2\fi)}%
    \def\citeauthoryear##1##2##3{##2}\@internalcite}
\def\shortciteANP{\def\@citeseppen{-1000}%
    \def\@cite##1##2{##1\if@tempswa , ##2\fi}%
    \def\citeauthoryear##1##2##3{##2}\@internalcite}
\def\citeyear{\def\@citeseppen{-1000}%
    \def\@cite##1##2{(##1\if@tempswa , ##2\fi)}%
    \def\citeauthoryear##1##2##3{##3}\@citedata}
\def\citeyearNP{\def\@citeseppen{-1000}%
    \def\@cite##1##2{##1\if@tempswa , ##2\fi}%
    \def\citeauthoryear##1##2##3{##3}\@citedata}
%
%
%
\def\@citedata{%
    \@ifnextchar [{\@tempswatrue\@citedatax}%
                  {\@tempswafalse\@citedatax[]}%
}

\def\@citedatax[#1]#2{%
\if@filesw\immediate\write\@auxout{\string\citation{#2}}\fi%
  \def\@citea{}\@cite{\@for\@citeb:=#2\do%
    {\@citea\def\@citea{, }\@ifundefined
       {b@\@citeb}{{\bf ?}%
       \@warning{Citation `\@citeb' on page \thepage \space undefined}}%
{\csname b@\@citeb\endcsname}}}{#1}}%

%
\def\@citex[#1]#2{%
\if@filesw\immediate\write\@auxout{\string\citation{#2}}\fi%
  \def\@citea{}\@cite{\@for\@citeb:=#2\do%
    {\@citea\def\@citea{; }\@ifundefined
       {b@\@citeb}{{\bf ?}%
       \@warning{Citation `\@citeb' on page \thepage \space undefined}}%
{\csname b@\@citeb\endcsname}}}{#1}}%

%
\def\@biblabel#1{}
\makeatother



\newdimen\bibindent
\bibindent=0.0em
\def\thebibliography#1{\section*{\refname}\list
   {}{\settowidth\labelwidth{[#1]}
   \leftmargin\parindent
   \itemindent -\parindent
   \listparindent \itemindent
   \itemsep 0pt
   \parsep 0pt}
   \def\newblock{}
   \sloppy
   \sfcode`\.=1000\relax}


\setlength{\baselineskip}{12.7pt}

\title{Simulating Misinformation Vulnerabilities With Agent Personas}

\author{\begin{center}David Farr\textsuperscript{1}, Lynnette Hui Xian Ng\textsuperscript{2}, Stephen Prochaska\textsuperscript{1},  Iain J. Cruickshank\textsuperscript{2}, and Jevin West\textsuperscript{1}\\
[11pt]
\textsuperscript{1}School of Information Science, University of Washington, Seattle, WA, USA\\
\textsuperscript{2}School of Computer Science, Carnegie Mellon University, Pittsburgh, PA, USA
\end{center}
}

\maketitle

\vspace{12pt}

\section*{ABSTRACT}
Disinformation campaigns can distort public perception and destabilize institutions. Understanding how different populations respond to information is crucial for designing effective interventions, yet real-world experimentation is impractical and ethically challenging. To address this, we develop an agent-based simulation using Large Language Models (LLMs) to model responses to misinformation. We construct agent personas spanning five professions and three mental schemas, and evaluate their reactions to news headlines. Our findings show that LLM-generated agents align closely with ground-truth labels and human predictions, supporting their use as proxies for studying information responses. We also find that mental schemas, more than professional background, influence how agents interpret misinformation. This work provides a validation of LLMs to be used as agents in an agent-based model of an information network for analyzing trust, polarization, and susceptibility to deceptive content in complex social systems.

\section{INTRODUCTION}
\label{sec:intro}

Protection against foreign information campaigns and the ability to conduct effective information operations are critical to modern national security. In an era where the information domain can be leveraged as a battlefield, there is a need to maintain information advantage, defined as ``the use, protection, and exploitation of information to achieve objectives more effectively than enemies and adversaries do" \shortcite{ADP3-13}. Achieving and sustaining information advantage requires not only the ability to disseminate compelling narratives but also to detect, counter, and mitigate adversarial information operations.

Foreign adversaries and non-state actors use information campaigns, which are sustained operations, to manipulate public perception, destabilize institutions, and degrade military readiness \shortcite{starbird2019disinformation,bradshaw2018challenging}. These campaigns often exploit cognitive biases, fracture public trust and shape operational environments before conflicts manifest kinetically \shortcite{ng2024exploring}. Although case studies of previous disinformation operations provide valuable information on population-based reactions \shortcite{reuter2019fake,tandoc2020diffusion}, the dynamic and adaptive nature of these operations presents a significant challenge for military planners and policymakers.

Real-world experimentation on populations is ethically and strategically untenable, making simulation-based approaches a critical alternative for research and operational planning with an added advantage of their ability to allow for exploration of diverse scenarios, parameters, and numerous trials \shortcite{epstein2008model}. Recent advancements in generative AI and agent-based modeling present new opportunities to study information operations in a controlled and scalable manner. By encoding AI-driven agents with distinct mental schemas, ideological frames, and cognitive biases, we can simulate how different populations perceive and react to competing narratives. Framing theory, which explores how individuals process and interpret information based on preexisting beliefs, provides a robust foundation for modeling adversarial messaging, population susceptibility, and counter-messaging strategies \shortcite{klein2007data}. LLM agents have been shown to be able to produce believable simulations of human interactions in social environments \shortcite{park2023generative,aher2023using}, and responses from LLM-based simulations have strong correlations with human subject experiments \shortcite{filippas2024large}. Integrating AI agents into disinformation simulations allows analysts to test information environment scenarios, evaluate disinformation resilience, and optimize civil-military engagement strategies.

This paper enables the systematic study of information competition by incorporating cognitive modeling techniques into Large Language Model (LLM)-based simulations. We use LLM-generated agents to simulate responses of different demographics towards misinformation. We build on previous work and use LLM-agents as a proxy to simulate responses from population groups. Through our simulation, we examine population-based reactions to misinformation, which enables better formulation and targeting of misinformation-combating strategies.



\section{RELATED WORK}

\subsection{Information Operations}
Information operations take many forms, ranging from bot networks amplifying simple messages to sophisticated, long-running campaigns that take advantage of pre-existing prejudices and biases to infiltrate online social networks and amplify divisions \shortcite{shao_spread_2018,arif_acting_2018,rid_active_2020}. Modern information operations are heavily participatory, in that publics engaging with strategically seeded content often believe that content to be true or reliable, leading them to amplify it and further spread false or misleading narratives \shortcite{starbird2019disinformation}. These “unwitting agents” are key to the spread of influence campaigns, making the boundaries of any given operation difficult to determine as the impacts of specific tactics have ripple effects outside of the direct control of strategists \shortcite{rid_active_2020}. Moreover, the content of campaigns often includes a mix of true, false, and misleading content to overwhelm audiences’ ability to make sense of novel or ambiguous events or information \shortcite{bittman_kgb_1985,rid_active_2020}.

In order to better identify and understand information operations, researchers have focused on different aspects of such operations, which are sometimes bucketed into three primary groups: actors, behaviors, and content \shortcite{francois_actors_2019}. Each of these categories is essential to any given operation, but for the current paper, we focus primarily on content and actors. One of the primary challenges facing the detection of strategic content is the difficulty in identifying the boundaries of an operation due to their targeting of pre-existing divisions within a target population \shortcite{bittman_kgb_1985,ellul_propaganda_1973,rid_active_2020}. In the process of targeting these divisions, operations often seek to impersonate or mimic people who fit particular stereotypes or expectations in order to appear more authentic \shortcite{arif_acting_2018}. This muddies the distinction between authentic and inauthentic activity such that detection has to rely on multiple signals simultaneously due to authentic and inauthentic content being very similar. 

This type of participation has become an integral part of modern influence campaigns, as strategists seek to guide or interrupt audiences as they seek to make sense of novel or ambiguous events \shortcite{starbird_influence_2023,starbird_what_is_going_on,prochaska_deep_storytelling}. Recent work has highlighted how different audiences interpret the same facts differently, allowing online influencers to opportunistically engage with and amplify strategic content and interpretive frames (which have also been referred to as schemas \shortcite{goffman_frame_1974,klein2007data}) that align with a specific audience’s expectations \shortcite{starbird_what_is_going_on}. We leverage this work, attempting to simulate members of particular audiences in order to better understand how those audiences might interpret the same facts differently. In order to do so, we take a broad definition of misinformation that includes the complex milieu of true, false, and misleading information that audiences would actually interact with were they to come into contact with an online information operation. Previous work has identified that common vectors for false or misleading information include unreliable media outlets and/or misleading headlines and stories \shortcite{grinberg_fake_2019} (see also \shortcite{bozarth_higher_2020}). Although campaigns often combine such headlines or stories with other tactics, we focus on this aspect of a campaign for our simulation as we are focused primarily on testing the ability of models to simulate diverse audience interpretations of misleading content.

\subsection{Simulation with AI agents}
Recent research has made notable strides in using language models as agents to simulate the spread of information on social systems. Agents constructed from generative AI models can simulate realistic human-like behavior and social interactions, and can therefore be used for modeling information flow \shortcite{park2023generative}. For example, LLMs can function as agents to assess the impact of network structure on the propagation of rumors \shortcite{hu2025simulating}. We extend this past work and compare the results of LLM-agents to human behavior, and incorporate real-world misinformation headlines, finding that LLM-agents can closely align with human perceptions of misinformation. 

Our agent-based modeling approach draws from social cognitive theory, which suggests that the information processing done by individual humans can be shaped by personal factors, behavioral patterns and environmental influences \shortcite{bandura2009social,ng2022pro}. In fact, individual cognitive differences can significantly predict susceptibility to fake news \shortcite{pennycook2019lazy}.  Building on this literature, our work incorporates human mental schemas that can affect an individual's susceptibility to misinformation, rather than simply focusing on adjusting the model's behavior through instructions.

Simulating more realistic agent characteristics, such as job titles and personality traits, can show emergence of news spread  and the effect of network topology on information dissemination \shortcite{li2024large}. Social media users exhibit variant misinformation sharing patterns. Researchers, for example, have shown that personality traits correlate with a propensity to share false content \shortcite{mosleh2021cognitive}. To align the human decision-making processes with real-world misinformation reaction, we build on prior work and simulate mental schemas that vary based on susceptibility to misinformation.

Further, role-based adjustments in ChatGPT prompts can impact the accuracy of misinformation detection \shortcite{haupt2024evaluating}, emphasizing the complexities involved in integrating biases and multiple perspectives into LLMs. Such experiments help us disentangle these complexities and provide insight into the nuanced ways in which biases in agent roles affect misinformation detection performance.

Much of the past research that uses LLMs as agents for simulating information spread focuses on network topology and prompt modifications. Some work indicate that LLMs can be effectively prompted to simulate diverse human perspectives on political discourse \shortcite{li2024political}, which allows the study of polarization dynamics across ideological spectrums. We bridge a research gap by simulating diverse humans to align LLM-generated agents with the mental schemas that humans have, an essential factor that is representative of real-world information environments. To do so, we simulate LLM-agents with mental schemas and professions, enabling a more interpretable comparison across different sets of headlines and messages.

\section{Methodology}

\begin{figure*}
    \centering
    \includegraphics[scale=.40]{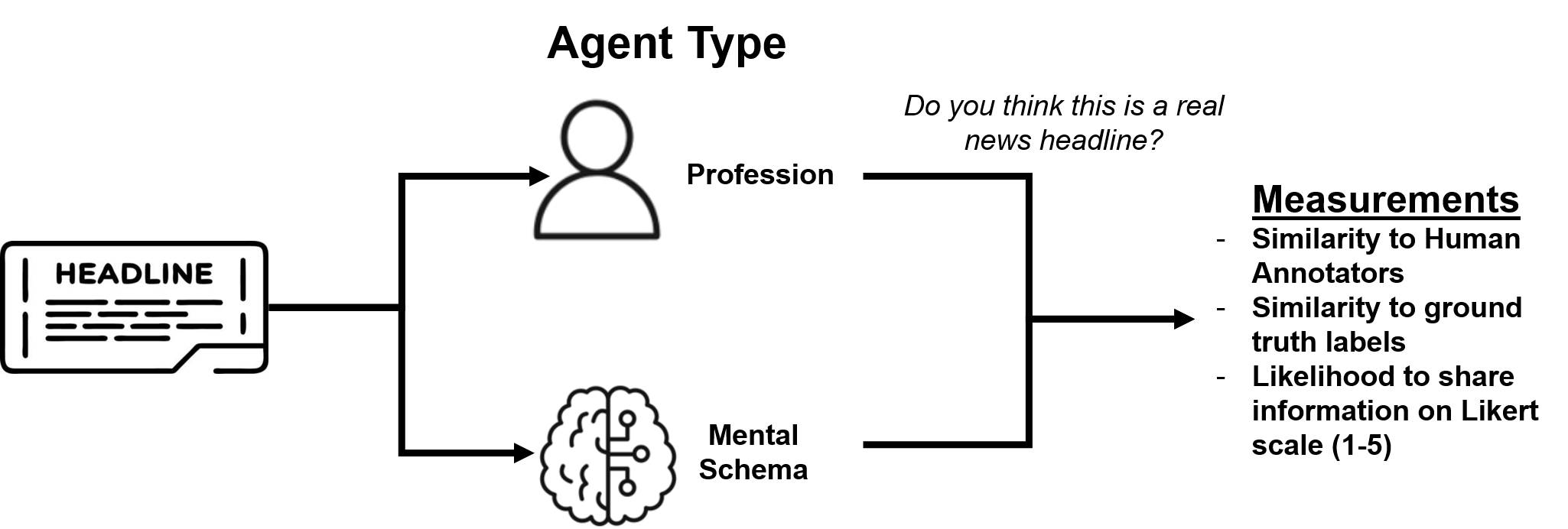}
    \caption{Overview of simulating reactions to misinformation by different agent personas.}
    \label{fig:sys_diagram}
\end{figure*}

\autoref{fig:sys_diagram} provides an overview of our simulation system. A headline from the Misinfo Reaction Frames corpus is read by LLM-Agent Personas that are constructed of different professions and mental schemas. These personas respond to the headline with their belief in the headline, and the likelihood the headline is a misinformation news. Their responses are compared with the gold labels and human predicted labels provided by the corpus as well as other agents.

\subsection{Agent Personas}
To ensure a diverse range of perspectives in our simulation, we designed eight distinct agent personas. We simulate these personas through LLMs. Each agent persona represents a specific professional background or a relevant mental schema. The selection of these agents was driven by their potential susceptibility or targeting in adversarial information campaigns. These personas are divided into two main groups: professions and mental schemas. 

Agent personas based on professions are: (1) \textit{military} personnel or soldiers, who are frequent targets of influence campaigns aimed at undermining morale, readiness, or battlefield decision-making \shortcite{gallacher2018junk}; \textit{college} students, a group that have a historical role in political activism and the early adoption of emerging narratives, and are a key demographic for grassroots and state-sponsored influence efforts \shortcite{levine1991undergraduates}; (3) \textit{retired} persons, representing the group of older adults that are frequently identified in misinformation research as a group that disproportionately engages and shares misleading content \shortcite{brashier2020aging}; (4) \textit{industrial} workers to present a blue-collar viewpoint; (5) \textit{financial} analysts for a contrasting white-collar viewpoint, allowing how disinformation targeting labor or socioeconomic groups resonate differently.

In addition to profession-based roles, we introduced agent personas based on mental schemas to model different cognitive responses to misinformation. These schemas are: (1) \textit{conspiracy}-believer that is highly receptive to conspiracy narratives and amplify fringe theories into mainstream discourse; (2) conspiracy-\textit{susceptible} agent that represents individuals who are not fully embedded in conspiracy thinking but remain prone to misinformation; (3) \textit{normal} persons who reads the news as a neutral baseline, allowing us to assess how an individual with no strong predispositions engages with the misinformation content. 

By incorporating this combination of occupational roles and cognitive frames, our agent-based simulation provides a robust framework for analyzing how different demographics process, propagate, or resist misinformation. We did not introduce explicit agent bias in prompting and instead relied on implicit model interpretation. This was a deliberate design choice that could be altered or studied in future work. This approach enables us to assess not only the effectiveness of disinformation campaigns but also possible intervention strategies to mitigate their impact in diverse populations. The full details of the prompts used for creating the agents are available in Appendix \ref{tab:agents_prompts}.

\subsection{Misinfo Reaction Frames corpus}  
We use the Misinfo Reaction Frames corpus to test agent reactions towards misinformation news \shortcite{misinfodata}. This corpus captures both the factual accuracy of news content and human cognitive responses to misinformation. The test dataset consists of 2,132 news headlines covering key domains of public discourse and national security concern, including COVID-19, climate change, and cancer misinformation. Each headline was fact-checked by researchers and assigned a binary classification as either misinformation or trustworthy information.  

Beyond factual classification, the dataset uniquely incorporates human reaction data, making it particularly suited for simulating real-world information environments, and therefore our simulation task. Each headline was evaluated by 63 human annotators recruited via Amazon Mechanical Turk, who provided annotations on:  

\begin{itemize}
    \item Perceived veracity – Whether they believed the headline to be real or fake.
    \item Emotional response – The dominant emotion elicited by the headline (e.g., fear, anger, trust).
    \item Propensity to share – A Likert-scale rating of how likely they would be to share the headline on social media.  
\end{itemize}

The cognitive and behavioral characteristics in the dataset that measures the perception of truth, emotional engagement, and likelihood of amplification, provide critical parameters for building agent-based simulations of digital influence operations. Unlike traditional misinformation datasets that focus solely on factual accuracy, this data set allows modeling of how different audiences respond to disinformation campaigns, which narratives are spread most effectively, and how misinformation resilience varies between demographic and ideological groups.  

By incorporating these human-centered response variables, our simulation can better approximate the complex social dynamics of digital information warfare than simulations purely based on alignment to gold labels or probability based susceptibility, offering insights into how adversaries exploit different population characteristics to spread disinformation.


\subsection{LLM-Based Simulation}
For each agent persona, we provided a headline from the Misinfo Reaction Frames Corpus dataset. We asked the agent two questions: (a) if the agent thinks the news headline is real, and (b) to rate the likelihood of the agent persona to share the information on a 1-5 Likert scale. We compare the results from (a) and (b) with human annotator predictions and ground truth labels from the original data.

In our experiments, we utilized the LLaMA 3.1 8B Instruct and GPT-4 models and compare the simulation performance across both models. GPT-4 is a larger reinforcement learning with human feedback (RLHF) model and LLaMA is a smaller, open-source model suitable for local execution. The two models represent different scales of model architecture.
Both models were run with a temperature setting of zero, which minimizes randomness in token selection and ensures more deterministic and reproducible outputs. Temperature controls the level of randomness in the model's predictions; a value of zero makes the model choose the highest-probability token at each step, rather than sampling from the full probability distribution. For GPT-4, a logit bias of 10 was applied to each token in our constrained set of labels. Logit bias allows us to artificially increase the likelihood of selecting specific tokens by adding a fixed value to their logits before the final softmax step, effectively steering the model toward those tokens. This adjustment helped optimize GPT-4's responses to the simulation task. All prompts are used are shown in Table \ref{tab:agents_prompts}.

\section{Results}

\subsection{GPT Simulation} 
\begin{figure*}[ht]
    \centering
    \includegraphics[scale=.4]{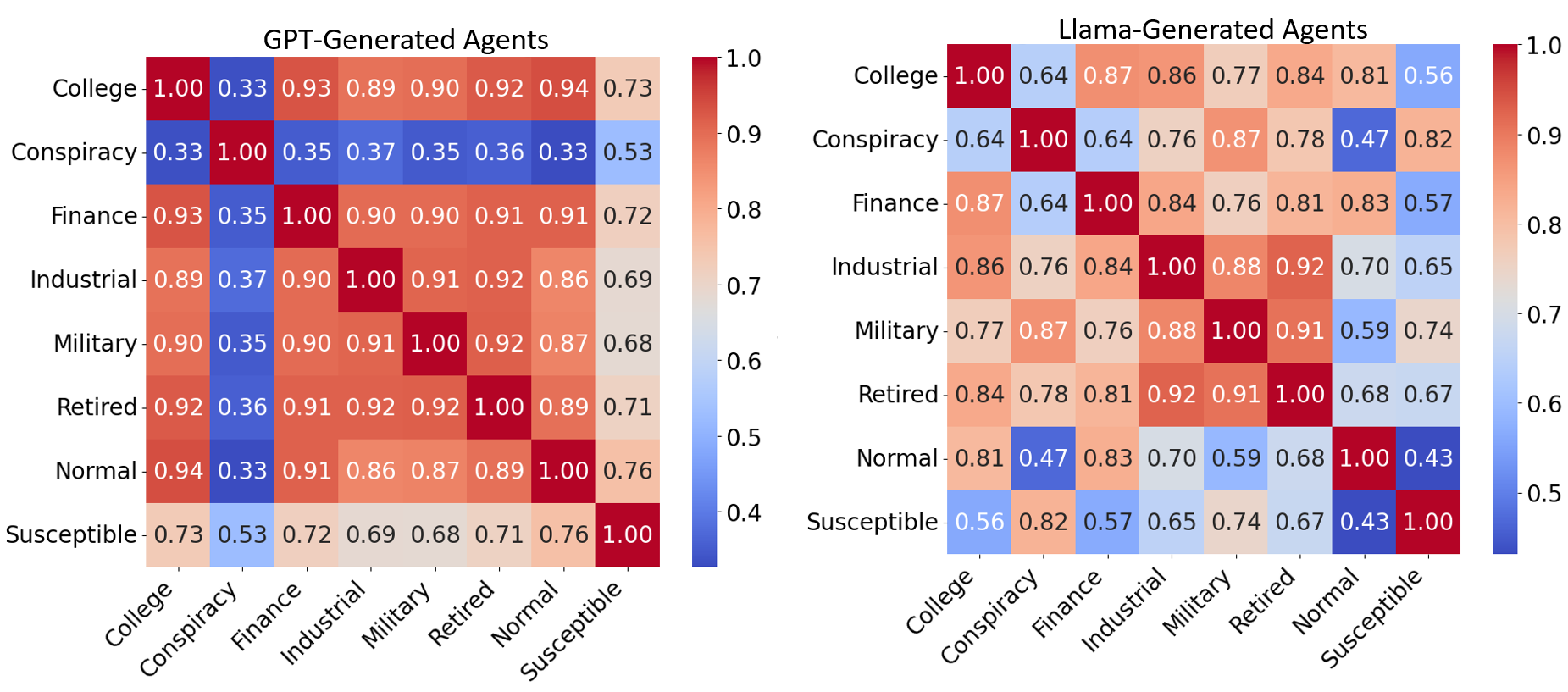}
    \caption{Heatmap of annotation agreement between LLM-generated agents on identifying whether a news headline constitutes misinformation.
}
    \label{fig:gpt_agent}
\end{figure*}

In our simulations using GPT, we consistently observe an ability of the agents to detect misinformation across different professional domains. This finding extrapolates to professions that can be assumed to differ significantly, such as financial workers and industrial workers, or groups such as young college students and military personnel. However, there were notable differences when agents were prompted with the various mental schemas, which consist of personas where agents were more susceptible to alternative news headlines or users that identify with conspiracy theories. As shown in the GPT-Generated Agents heatmap in Figure \ref{fig:gpt_agent} there is little agreement (0.53)  between agents that were \textit{conspiracy} and \textit{susceptible}, and even less agreement (0.33) between agents that were \textit{conspiracy} and \textit{normal}. These observations suggests that the different mental states affect the difference in response to misinformation more than professions do.

One interesting observation was that prompting GPT as a neutral news reader (\textit{normal}) leads to the highest overlap in annotations with human annotators, resulting in the best performance on the task of identifying misinformation. This suggests that a more neutral, unbiased approach may better align with human decision-making processes in misinformation detection and that GPT is relatively well aligned with human perceptions of misinformation.

Out of the eight GPT generated agent annotators, six outperformed human annotators in identifying misinformation, achieving over 63\% accuracy (Figure \ref{fig:gold}). However, as shown in Figure \ref{fig:gpt_agent} under GPT-Generated Agents, there were significant differences in agreement between agent annotators. In particular, the conspiracy-driven and susceptible agents demonstrate a stronger tendency to classify misinformation as true. Interestingly, even among agents who only differed in profession (e.g., financial workers versus industrial workers), there were notable disagreements in their assessments of information truthfulness. Although the overall precision of the identification of misinformation was similar, the agreement on individual data points had a difference of 10\% (Figure \ref{fig:gpt_agent}).

\subsection{LLaMA Simulation} 
\begin{figure*}[ht]
    \centering
    \includegraphics[scale=.43]{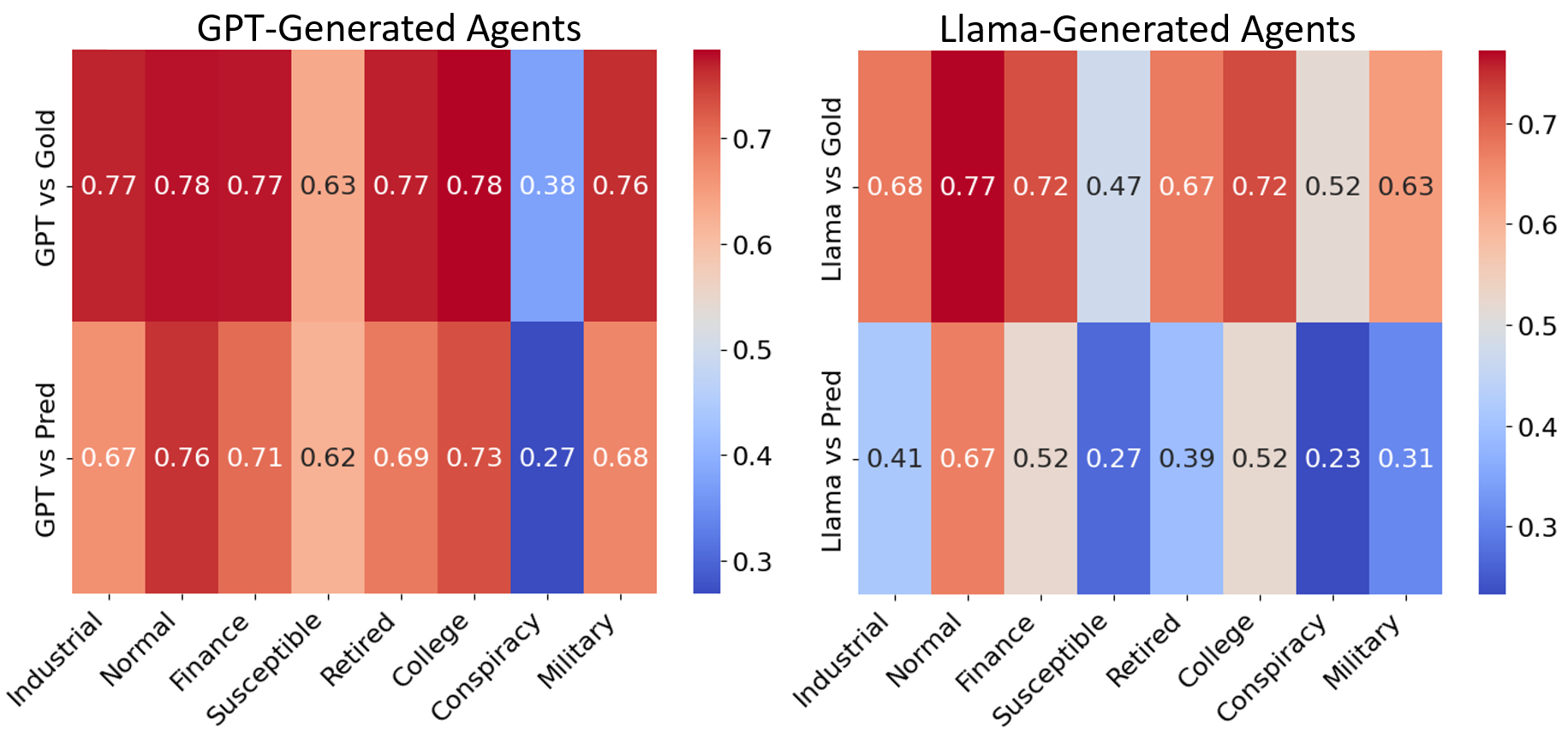}
    \caption{Comparison of LLM-generated agent predictions to gold labels and human annotator judgments. LLM Model versus gold shows the comparison of each individual LLM-based agent to gold annotations and GPT vs Pred shows the comparison of each individual LLM-based agent to human predictions.}
    \label{fig:gold}
\end{figure*}

In contrast to GPT, the simulation using the LLaMA model for agent persona generation exhibited much greater variance in performance. As with GPT, the neutral news reader (\textit{normal}) agent yielded the highest classification performance and most closely mirrored the human annotations. As shown in the LlaMa-Generated agent heatmap in Figure \ref{fig:gold}, five of the eight LLaMA agents outperformed human annotators in identifying misinformation (more than 63\% accuracy), although their performance was worse than that of the GPT agents.

Additionally, the LLaMA agents did not align with human annotations as frequently as GPT agents did, suggesting that LLaMA’s outputs are less consistent in terms of reflecting human annotation judgment. The observed variance in responses among the LLaMA agents could be interpreted as a useful representation of the diversity of perspectives that exist among individuals in the information environment, but additional fine-tuning should be done in future work to better align agent behavior with human interpretation.

We compare the similarity of the responses of whether the LLM-generated agents think that the input headline is real in two forms: the LLM-generated agents within each LLM-model, and the LLM-generated responses against the gold labels and human annotator judgments. 

\autoref{fig:gpt_agent} shows the agreement heatmap of LLM-generated agents towards whether a news headline constitutes misinformation. LLaMA-generated agents exhibit greater variance than GPT-generated agents outcomes across both professions and mental schemas. Agents assigned different professions exhibited largely similar interpretations of information (e.g., finance vs college), whereas altering the agents' mental schemas (e.g. conspiracy vs normal) led to significant variations in their classifications of misinformation versus real information.

\autoref{fig:gold} compares the LLM-generated agent predictions to gold labels and human annotator judgments. The gold labels and human judgments are provided in the original Misinfo Reaction Corpus. GPT-generated agents align more closely with gold labels than with human annotators and demonstrate higher accuracy in identifying misinformation. The agents' similarity to human predictions suggest that GPT-generated agents can serve as effective proxies to simulate responses to misinformation. On the other hand, while LLaMA-generated agents effectively identify misinformation based on gold labels, they exhibited limited alignment with human predictions. Additional tuning might be necessary to enhance their alignment with human perception, especially when using role prompting to understand responses of diverse perspectives.

\subsection{Propensity to Share} 
\begin{figure*}[ht] \centering \includegraphics[scale=.65]{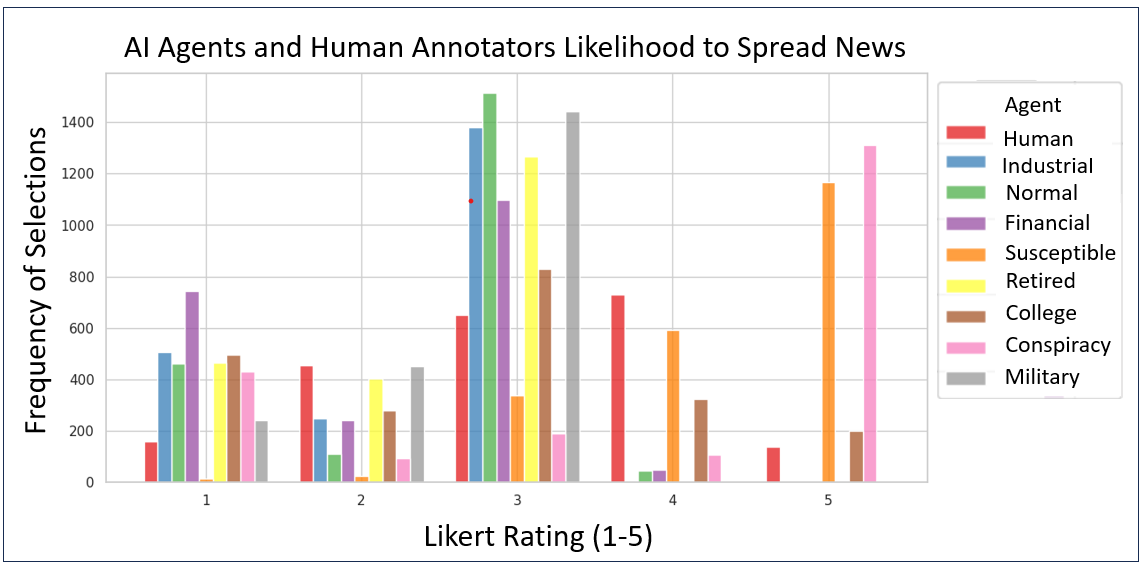} \caption{Bar plot comparing Likert scale ratings of the likelihood to share news headlines, as assessed by LLM agents and human annotators with 1 being not at all likely and five being extremely likely. Agents tend to cluster toward the middle of the Likert scale where humans seem to be more evenly distributed.} 
\label{fig:likert} 
\end{figure*}

\autoref{fig:likert} compares the Likert scale ratings of the likelihood to share news headlines between GPT-based agents and human annotators. Although most AI agents show general alignment with human ratings, agents with significant schema adjustments, such as susceptible and conspiracy agents, exhibit notable deviations, showing a higher propensity to share information, including misinformation. These agents tend to cluster their sharing likelihood into common response categories, regardless of the specific headline. This contrasts with human annotators, whose Likert ratings are more evenly distributed across the scale, though they still most often choose a medium likelihood rating (i.e., choice three of five). For a more nuanced comparison, human annotators would need to be categorized using the same schema as the agents to assess whether agents accurately mimic their representative human counterparts. Nonetheless, when averaging across all agents and Likert responses, agent behavior appears to resemble overall human patterns.

\section{Discussion}
\label{sec:discussion}
Our findings indicate that LLM-simulated agents largely align with both ground truth labels and human-annotated predictions in their interpretations of misinformation. However, most agent types deviated from human norms when it came to sharing behavior. While human annotators exhibited a relatively even distribution across the Likert scale for likelihood of sharing, simulated agents tended to cluster around a moderate likelihood. The type of language model used to simulate agents significantly affected the variance in misinformation labeling and the degree of agreement both among agents and between agents and humans. Notably, GPT-based agents exhibited lower inter-agent variance and greater alignment with human judgments, which we interpret as a strength, reflecting the diversity and complexity of real-world information environments through variance in interpretation while preserving alignment. These results suggest that LLM-simulated agents can serve as effective proxies for analyzing human responses to misinformation. \autoref{tab:examples} provides sample responses indicating whether the simulated agent persona judged a headline as real or as misinformation.

\begin{table}
\centering
\caption{GPT-Generated Agent responses towards ``Do you think this is a real news headline?" question. \textbf{bold} items means the responses agree with the Gold labels. This table was designed to show examples where there was large disagreement among agents and is not representative of agent accuracy, as shown by the conspiracy agent correctly classifying many difficult headlines.}
\small
\begin{tabular}{|p{2.5cm}|*{5}{>{\raggedright\arraybackslash}p{2.3cm}|}}
\hline
\textbf{Headline/Agent Category} & Ukrainian schools will be closed as lockdown measure & AP: Five years on, Paris climate accord working; CO2 emissions dropped 17\% this year & Siberian Environmentalist Detained for Poaching World's Largest Sheep & How climate change could benefit Russia & CBS: Too Many 'Thank You' Emails Contribute to Global Warming \\
\hline
\textbf{Gold} & misinfo & misinfo & real & real & misinfo \\
\hline
\textbf{Predicted} & real & real & \textbf{real} & misinfo & \textbf{misinfo} \\
\hline
\textbf{College} & \textbf{misinfo} & real & misinfo & misinfo & \textbf{misinfo} \\
\hline
\textbf{Industrial} & \textbf{misinfo} & real & misinfo & misinfo & \textbf{misinfo} \\
\hline
\textbf{Financial} & real & real & misinfo & misinfo & \textbf{misinfo} \\
\hline
\textbf{Retired} & real & real & misinfo & misinfo & \textbf{misinfo} \\
\hline
\textbf{Military} & real & real & misinfo & misinfo & \textbf{misinfo} \\
\hline
\textbf{Standard} & real & real & misinfo & \textbf{real} & \textbf{misinfo} \\
\hline
\textbf{Susceptible} & real & \textbf{misinfo} & misinfo & \textbf{real} & real \\
\hline
\textbf{Conspiracy} & \textbf{misinfo} & \textbf{misinfo} & \textbf{real} & \textbf{real} & real \\
\hline
\end{tabular}
\label{tab:examples}
\end{table}

A key finding of our experiments is that an agent’s interpretation of information was more strongly influenced by mental schemas than by professional background. While responses were relatively consistent across different professions, significant differences emerged when agents were prompted with varying cognitive predispositions, such as susceptibility to alternative news sources or belief in conspiracy theories. In fact, compared to the \textit{normal} agent persona, the \textit{conspiracy} and \textit{susceptible} persona performs worse in detecting misinformation, showing that pre-existing beliefs can distort a person's information processing. This is in line with the phenomenon of confirmation bias or motivated reasoning. This insight underscores the importance of tailoring misinformation intervention strategies to cognitive and ideological predispositions rather than professional affiliations. This finding also has practical implications for misinformation mitigation efforts. Rather than designing interventions that target individuals based on profession, strategies may want to focus on addressing the underlying cognitive biases that contribute to susceptibility.

From a methodological perspective, our work establishes a pipeline for systematically creating agent personas and simulating their responses towards misinformation news headlines. This simulation framework creates a controlled environment that allows isolation of specific cognitive and profession-based characteristics that might affect susceptibility to misinformation. The framework also allows for rapid iteration and large scale testing across population slices that will be difficult to implement with human studies like surveys.

Beyond misinformation detection, our approach highlights the potential of LLM-based agent simulations for studying broader information dynamics. The ability to model and test different personas enables researchers to analyze how different demographic and cognitive groups interact with and propagate information in digital spaces. This technique could extend to applications such as assessing polarization, trust in media, and susceptibility to other forms of deceptive content, including deepfakes and AI-generated propaganda. Further, from a national security perspective, understanding the susceptibility towards misinformation belief and sharing of population segments (i.e., profession, mental schemas) allows for more precise design of targeted counter-disinformation efforts such as educational materials. Specific interventions may need to be tailored towards each mental schema for better effectiveness. 


\paragraph{Limitations and Future Work} Our study only investigated two LLMs, and while these two models indicated good alignment with the gold label and human annotations, future work can investigate the construction of agent personas with different LLMs, even comparing the similarity and differences between the agent personas constructed from different LLMs. Further, a larger range of professions and mental schemas can be incorporated into the agent persona construction to showcase a wider diversity of personas. Nonetheless, this work offers a preliminary investigation into measuring the responses towards misinformation of different personas and their alignment with human predictions.

\section{Conclusion}
In this study, we constructed multiple agent personas using Large Language Models (LLMs) and evaluated their responses to misinformation headlines. Our results indicate that LLM-simulated agents can effectively approximate human decision-making patterns in misinformation detection, with a notable variance in susceptibility based on cognitive predisposition rather than professional backgrounds. 

Our study highlights the value of LLM-based agents as a tool for understanding the mechanisms of misinformation spread. By simulating diverse personas, researchers can systematically test intervention strategies, study information propagation dynamics, and assess the effectiveness of counter-misinformation efforts at scale.

As the digital information landscape continues to evolve, the ability to simulate complex human-like responses with multi-agent frameworks presents opportunities to create robust disinformation defense strategies through a more realistic information environment than was previously possible. Future works should explore additional model architectures, fine-tuning methods, and real-world behavioral comparisons to refine accuracy and applicability of these simulations. Integrating LLM-based agent modeling into misinformation research can provide insights into how misinformation spreads, what populations are particularly vulnerable to specific messaging types, and how to counteract its effects in an increasingly AI driven information environment.

\appendix

\section{APPENDICES} 

\subsection{Agent Prompts} 
\autoref{tab:agents_prompts} lists the agent prompts used for each profession and mental schema. The design decision was made to not impose explicit biases in the prompt of profession traits and instead let the agent impose those traits itself.

\begin{table}[ht]
    \centering
    \caption{Referenced agents and their system prompts.}
    \label{tab:agents_prompts}
    \begin{tabular}{|l|p{10cm}|}
        \hline
        \textbf{Agent Name} & \textbf{System Prompt} \\ 
        \hline
        Soldier & "You are a young enlisted soldier and read the following headline: HEADLINE Did you believe the headline?"" \\ 
        \hline
        College Student & "You are a politically active college student and read the following headline: HEADLINE Did you believe the headline?" \\ 
        \hline
        Retired & "You are an older retired professional and read the following headline: HEADLINE Did you believe the headline?"" \\ 
        \hline
        Industrial Worker & "You are an industrial worker and read the following headline: HEADLINE Did you believe the headline?" \\ 
        \hline
        Financial Analyst & "You are a financial analyst and read the following headline: HEADLINE Did you believe the headline?" \\ 
        \hline
        Conspiracy-Believer & "You are an individual who believes in many conspiracy theories and read the following headline: HEADLINE Did you believe the headline?"" \\ 
        \hline
        Conspiracy-Susceptible & "You are someone who is susceptible to conspiracy theories and read the following headline: HEADLINE Did you believe the headline?"" \\ 
        \hline
        Standard News Reader & "You are a standard news reader and read the following headline: HEADLINE Did you believe the headline?""\\ 
        \hline
    \end{tabular}
\end{table}

\footnotesize

\bibliographystyle{wsc}

\bibliography{references}

\begin{thebibliography}{}

\bibitem[\protect\citeauthoryear{Aher, Arriaga, and Kalai}{Aher et~al.}{2023}]{aher2023using}
Aher, G.~V., R.~I. Arriaga, and A.~T. Kalai. 2023.
\newblock ``Using Large Language Models to Simulate Multiple Humans and Replicate Human Subject Studies''.
\newblock In {\em Proceedings of the 40th International Conference on Machine Learning}, edited by\ A.~Krause, E.~Brunskill, K.~Cho, B.~Engelhardt, S.~Sabato, and J.~Scarlett, Volume 202 of {\em Proceedings of Machine Learning Research}.
\newblock July 23\textsuperscript{rd}-29\textsuperscript{th}, Honolulu, Hawaii, 337-371.

\bibitem[\protect\citeauthoryear{Arif, Stewart, and Starbird}{Arif et~al.}{2018}]{arif_acting_2018}
Arif, A., L.~G. Stewart, and K.~Starbird. 2018.
\newblock ``Acting the Part: Examining Information Operations Within \#BlackLivesMatter Discourse''.
\newblock In {\em Proceedings of the ACM on Human-Computer Interaction}, Volume~2.
\newblock New York, NY, USA: November 3\textsuperscript{rd}-7\textsuperscript{th}, Jersey City, New Jersey, 20:1-20:27.

\bibitem[\protect\citeauthoryear{Bandura}{Bandura}{2009}]{bandura2009social}
Bandura, A. 2009.
\newblock ``Social Cognitive Theory of Mass Communication''.
\newblock {\em Media Psychology\/}:265--299.


\bibitem[\protect\citeauthoryear{Bittman}{Bittman}{1985}]{bittman_kgb_1985}
Bittman, L. 1985, October.
\newblock {\em The {KGB} and {Soviet} {Disinformation}: {An} {Insider}'s {View}\/}. 1st Edition ed.
\newblock Washington: Pergamon Pr.


\bibitem[\protect\citeauthoryear{Bozarth, Saraf, and Budak}{Bozarth et~al.}{2020}]{bozarth_higher_2020}
Bozarth, L., A.~Saraf, and C.~Budak. 2020.
\newblock ``Higher Ground? How Groundtruth Labeling Impacts Our Understanding of Fake News about the 2016 US Presidential Nominees''.
\newblock In {\em Proceedings of the International AAAI Conference on Web and Social Media}, Volume~14.
\newblock February 7\textsuperscript{th}-12\textsuperscript{th}, New York, NY, 48-59.

\bibitem[\protect\citeauthoryear{Bradshaw and Howard}{Bradshaw and Howard}{2018}]{bradshaw2018challenging}
Bradshaw, S., and P.~N. Howard. 2018.
\newblock ``Challenging Truth and Trust: A Global Inventory of Organized Social Media Manipulation''.
\newblock {\em The Computational Propaganda Project\/}~1:1--26.


\bibitem[\protect\citeauthoryear{Brashier and Schacter}{Brashier and Schacter}{2020}]{brashier2020aging}
Brashier, N.~M., and D.~L. Schacter. 2020.
\newblock ``Aging in an Era of Fake News''.
\newblock {\em Current Directions in Psychological Science\/}~29(3):316--323.


\bibitem[\protect\citeauthoryear{Ellul}{Ellul}{1973}]{ellul_propaganda_1973}
Ellul, J. 1973, January.
\newblock {\em Propaganda: {The} {Formation} of {Men}'s {Attitudes} {\textbar} {Politics} and {Prose} {Bookstore}}.
\newblock ISBN: 9780394718743.


\bibitem[\protect\citeauthoryear{Epstein}{Epstein}{2008}]{epstein2008model}
Epstein, J.~M. 2008.
\newblock ``Why Model?''.
\newblock {\em Journal of Artificial Societies and Social Simulation\/}~11(4):12.


\bibitem[\protect\citeauthoryear{Filippas, Horton, and Manning}{Filippas et~al.}{2024}]{filippas2024large}
Filippas, A., J.~J. Horton, and B.~S. Manning. 2024.
\newblock ``Large Language Models as Simulated Economic Agents: What Can We Learn from Homo Silicus?''.
\newblock In {\em Proceedings of the 25th ACM Conference on Economics and Computation}.
\newblock July 8\textsuperscript{th}-11\textsuperscript{th}, New Haven, CT, 614-615.

\bibitem[\protect\citeauthoryear{Fran{\c{c}}ois}{Fran{\c{c}}ois}{2019}]{francois_actors_2019}
Fran{\c{c}}ois, C. 2019.
\newblock ``Actors, Behaviours, Content: A disinformation ABC (Transatlantic High Level Working Group on Content Moderation Online and Freedom of Expression Series). Annenberg Public Policy Center, University of Pennsylvania; Annenberg Foundation Trust, Sunnylands''.
\newblock {\em Institute for Information Law, University of Amsterdam\/}.


\bibitem[\protect\citeauthoryear{Gabriel, Hallinan, Sap, Nguyen, Roesner, Choi, and Choi}{Gabriel et~al.}{2022}]{misinfodata}
Gabriel, S., S.~Hallinan, M.~Sap, P.~Nguyen, F.~Roesner, E.~Choi {\em et~al}. 2022, May.
\newblock ``Misinfo Reaction Frames: Reasoning about Readers' Reactions to News Headlines''.
\newblock In {\em Proceedings of the 60th Annual Meeting of the Association for Computational Linguistics (Volume 1: Long Papers)}, edited by\ S.~Muresan, P.~Nakov, and A.~Villavicencio.
\newblock May 22\textsuperscript{nd}-27\textsuperscript{th}, Dublin, Ireland, 3108-3127: Association for Computational Linguistics~\url{https://doi.org/10.18653/v1/2022.acl-long.222}.

\bibitem[\protect\citeauthoryear{Gallacher, Barash, Howard, and Kelly}{Gallacher, John D and Barash, Vlad and Howard, Philip N and Kelly, John}{2018}]{gallacher2018junk}
Gallacher, John D and Barash, Vlad and Howard, Philip N and Kelly, John 2018.
\newblock ``Junk News on Military Affairs and National Security: Social Media Disinformation Campaigns against US Military Personnel and Veterans''.
\newblock Last modified Feb 10, 2018. \url{https://arxiv.org/abs/1802.03572}.

\bibitem[\protect\citeauthoryear{Goffman}{Goffman}{1974}]{goffman_frame_1974}
Goffman, E. 1974.
\newblock {\em Frame analysis: {An} Essay on the Organization of Experience}.
\newblock Frame analysis: {An} Essay on the Organization of Experience. Cambridge, MA, US: Harvard University Press.
\newblock Pages: ix, 586.


\bibitem[\protect\citeauthoryear{Grinberg, Joseph, Friedland, Swire-Thompson, and Lazer}{Grinberg et~al.}{2019}]{grinberg_fake_2019}
Grinberg, N., K.~Joseph, L.~Friedland, B.~Swire-Thompson, and D.~Lazer. 2019, January.
\newblock ``Fake News on {Twitter} During the 2016 {U}.{S}. Presidential Election''.
\newblock {\em Science\/}~363(6425):374--378.


\bibitem[\protect\citeauthoryear{Haupt, Yang, Purnat, and Mackey}{Haupt et~al.}{2024}]{haupt2024evaluating}
Haupt, M.~R., L.~Yang, T.~Purnat, and T.~Mackey. 2024.
\newblock ``Evaluating the Influence of Role-Playing Prompts on ChatGPT’s Misinformation Detection Accuracy: Quantitative Study''.
\newblock {\em JMIR infodemiology\/}~4(1):e60678.


\bibitem[\protect\citeauthoryear{Hu, Liakopoulos, Wei, Marculescu, and Yadwadkar}{Hu, Tianrui and Liakopoulos, Dimitrios and Wei, Xiwen and Marculescu, Radu and Yadwadkar, Neeraja J}{2025}]{hu2025simulating}
Hu, Tianrui and Liakopoulos, Dimitrios and Wei, Xiwen and Marculescu, Radu and Yadwadkar, Neeraja J 2025.
\newblock ``Simulating Rumor Spreading in Social Networks using LLM Agents''.
\newblock Last modified Feb 3, 2025. \url{https://arxiv.org/abs/2502.01450}.

\bibitem[\protect\citeauthoryear{Klein, Phillips, Rall, and Peluso}{Klein et~al.}{2007}]{klein2007data}
Klein, G., J.~K. Phillips, E.~L. Rall, and D.~A. Peluso. 2007.
\newblock ``A Data--Frame Theory of Sensemaking''.
\newblock In {\em Expertise out of context}, edited by\ R.~R. Hoffman,  118--160. London: Psychology Press.

\bibitem[\protect\citeauthoryear{Levine and Hirsch}{Levine and Hirsch}{1991}]{levine1991undergraduates}
Levine, A., and D.~Hirsch. 1991.
\newblock ``Undergraduates in Transition: A New Wave of Activism on American College Campus''.
\newblock {\em Higher Education\/}~22(2):119--128.


\bibitem[\protect\citeauthoryear{Li, Li, Chen, Gui, Yang, Yu, Wang, Cai, Zhou, Shen, et~al.}{Li, Lincan and Li, Jiaqi and Chen, Catherine and Gui, Fred and Yang, Hongjia and Yu, Chenxiao and Wang, Zhengguang and Cai, Jianing and Zhou, Junlong Aaron and Shen, Bolin and others}{2025}]{li2024political}
Li, Lincan and Li, Jiaqi and Chen, Catherine and Gui, Fred and Yang, Hongjia and Yu, Chenxiao and Wang, Zhengguang and Cai, Jianing and Zhou, Junlong Aaron and Shen, Bolin and others 2025.
\newblock ``Political-llm: Large Language Models in Political Science''.
\newblock Last modified Dec 9, 2024. \url{https://arxiv.org/abs/2412.06864}.

\bibitem[\protect\citeauthoryear{Li, Xu, Zhang, and Malthouse}{Li, Xinyi and Xu, Yu and Zhang, Yongfeng and Malthouse, Edward C}{2024}]{li2024large}
Li, Xinyi and Xu, Yu and Zhang, Yongfeng and Malthouse, Edward C 2024.
\newblock ``Large Language Model-Driven Multi-Agent Simulation for News Diffusion Under Different Network Structures''.
\newblock Last modified Oct 16, 2024. \url{https://arxiv.org/abs/2410.13909}.

\bibitem[\protect\citeauthoryear{Mosleh, Pennycook, Arechar, and Rand}{Mosleh et~al.}{2021}]{mosleh2021cognitive}
Mosleh, M., G.~Pennycook, A.~A. Arechar, and D.~G. Rand. 2021.
\newblock ``Cognitive Reflection Correlates with Behavior on Twitter''.
\newblock {\em Nature Communications\/}~12(1):921.


\bibitem[\protect\citeauthoryear{Ng and Carley}{Ng and Carley}{2022}]{ng2022pro}
Ng, L. H.~X., and K.~M. Carley. 2022.
\newblock ``Pro or Anti? A Social Influence Model of Online Stance Flipping''.
\newblock {\em IEEE Transactions on Network Science and Engineering\/}~10(1):3--19.


\bibitem[\protect\citeauthoryear{Ng, Zhou, and Carley}{Ng, Lynnette Hui Xian and Zhou, Wenqi and Carley, Kathleen M}{2024}]{ng2024exploring}
Ng, Lynnette Hui Xian and Zhou, Wenqi and Carley, Kathleen M 2024.
\newblock ``Exploring Cognitive Bias Triggers in Covid-19 Misinformation Tweets: A Bot vs. Human Perspective''.
\newblock Last modified June 11, 2024. \url{https://arxiv.org/abs/2406.07293}.

\bibitem[\protect\citeauthoryear{Park, O'Brien, Cai, Morris, Liang, and Bernstein}{Park et~al.}{2023}]{park2023generative}
Park, J.~S., J.~O'Brien, C.~J. Cai, M.~R. Morris, P.~Liang, and M.~S. Bernstein. 2023.
\newblock ``Generative Agents: Interactive Simulacra of Human Behavior''.
\newblock In {\em Proceedings of the 36th Annual ACM Symposium on User Interface Software and Technology}.
\newblock October 29\textsuperscript{th}-Novemeber 1\textsuperscript{st}, San Fransisco, California, 1-22.

\bibitem[\protect\citeauthoryear{Pennycook and Rand}{Pennycook and Rand}{2019}]{pennycook2019lazy}
Pennycook, G., and D.~G. Rand. 2019.
\newblock ``Lazy, Not Biased: Susceptibility to Partisan Fake News is Better Explained by Lack of Reasoning Than by Motivated Reasoning''.
\newblock {\em Cognition\/}~188:39--50.


\bibitem[\protect\citeauthoryear{Prochaska, Vera, Lew~Tan, Yamron, Venuto, Kejriwal, Chu, and Starbird}{Prochaska et~al.}{2025}]{prochaska_deep_storytelling}
Prochaska, S., J.~Vera, D.~Lew~Tan, B.~Yamron, S.~Venuto, A.~Kejriwal,  {\em et~al}. 2025.
\newblock ``Deep {Storytelling}: {Collective} {Sensemaking} and {Layers} of {Meaning} in {U}.{S}. {Elections}''.
\newblock Number CSCW1.
\newblock Submitted for publication to \emph{Conference on {Computer}-{Supported} {Cooperative} {Work} \& {Social} {Computing}} October 18\textsuperscript{th}-22\textsuperscript{nd}, Bergen, Norway.

\bibitem[\protect\citeauthoryear{Reuter, Hartwig, Kirchner, and Schlegel}{Reuter et~al.}{2019}]{reuter2019fake}
Reuter, C., K.~Hartwig, J.~Kirchner, and N.~Schlegel. 2019.
\newblock ``Fake News Perception in Germany: A Representative Study of People's Attitudes and Approaches to Counteract Disinformation''.
\newblock {\em Association for Information Systems AIS\/}.


\bibitem[\protect\citeauthoryear{Rid}{Rid}{2020}]{rid_active_2020}
Rid, T. 2020, April.
\newblock {\em Active Measures : The Secret History of Disinformation and Political Warfare\/}. First edition. ed.
\newblock New York: Farrar, Straus and Giroux.
\newblock ISBN: 9780374287269.


\bibitem[\protect\citeauthoryear{Shao, Ciampaglia, Varol, Yang, Flammini, and Menczer}{Shao et~al.}{2018}]{shao_spread_2018}
Shao, C., G.~L. Ciampaglia, O.~Varol, K.-C. Yang, A.~Flammini, and F.~Menczer. 2018, November.
\newblock ``The Spread of Low-Credibility Content by Social Bots''.
\newblock {\em Nature Communications\/}~9(1):4787~\url{https://doi.org/10.1038/s41467-018-06930-7}.


\bibitem[\protect\citeauthoryear{Starbird, Arif, and Wilson}{Starbird et~al.}{2019}]{starbird2019disinformation}
Starbird, K., A.~Arif, and T.~Wilson. 2019.
\newblock ``Disinformation as Collaborative Work: Surfacing the Participatory Nature of Strategic Information Operations''.
\newblock In {\em Proceedings of the ACM on Human-Computer Interaction}, Volume~3.
\newblock November 9\textsuperscript{th}-13\textsuperscript{th}, Austin, TX, 1-26.

\bibitem[\protect\citeauthoryear{Starbird, DiResta, and DeButts}{Starbird et~al.}{2023}]{starbird_influence_2023}
Starbird, K., R.~DiResta, and M.~DeButts. 2023, June.
\newblock ``Influence and {Improvisation}: {Participatory} {Disinformation} during the 2020 {US} {Election}''.
\newblock {\em Social Media + Society\/}.


\bibitem[\protect\citeauthoryear{Starbird, Prochaska, and Yamron}{Starbird et~al.}{2025}]{starbird_what_is_going_on}
Starbird, K., S.~Prochaska, and B.~Yamron. 2025.
\newblock ``What is Going On? An Evidence-Frame Framework for Analyzing Online Rumors About Election Integrity''.
\newblock Submitted for publication to \emph{Conference on {Computer}-{Supported} {Cooperative} {Work} \& {Social} {Computing}} October 18\textsuperscript{th}-22\textsuperscript{nd}, Bergen, Norway.

\bibitem[\protect\citeauthoryear{Tandoc~Jr, Lim, and Ling}{Tandoc~Jr et~al.}{2020}]{tandoc2020diffusion}
Tandoc~Jr, E.~C., D.~Lim, and R.~Ling. 2020.
\newblock ``Diffusion of Disinformation: How Social Media Users Respond to Fake News and Why''.
\newblock {\em Journalism\/}~21(3):381--398.


\bibitem[\protect\citeauthoryear{{U.S. Department of the Army}}{{U.S. Department of the Army}}{2023}]{ADP3-13}
{U.S. Department of the Army} 2023.
\newblock {\em {Army Doctrine Publication (ADP) 3-13: Information Operations}}.
\newblock Washington, D.C.: Headquarters, Department of the Army.

\end{thebibliography}

\section*{AUTHOR BIOGRAPHIES}
\noindent {\bf \MakeUppercase{David Farr}} is a PhD student at the University of Washington in the School of Information Science and a co-first author of this work. He received his MSc in Operational Research with Data Science from the University of Edinburgh, and he did his Bachelor's degree in Systems Engineering from the United States Military Academy at West Point. His research interests include multi-agent systems, network analysis, and data annotation. His e-mail address is \email{dtfarr@uw.edu} and his website is \url{https://davidthfarr.github.io/}. \\

\noindent {\bf \MakeUppercase{Lynnette Hui Xian Ng}} is a PhD student in Societal Computing at Carnegie Mellon University, School of Computer Science and a co-first author of this work. She did her bachelor's degree in Computer Science at National University of Singapore. Her current work focuses around network and engagement impact of behavioral influence techniques applied by automated agents on social media. Her e-mail address is \email{lynnetteng@cmu.edu} and his website is \url{https://quarbby.github.io}.\\

\noindent {\bf \MakeUppercase{Stephen Prochaska}} is a PhD candidate in the Information Science Department at the University of Washington. His current work focuses around sensemaking, framing theory, and information campaigns. His e-mail address is \email{sprochas@uw.edu}.\\

\noindent {\bf \MakeUppercase{Iain J. Cruickshank}} is adjunct faculty at Carnegie-Mellon University's Software and Societal Systems Department and Lecturer at Johns Hopkins University. His research interests include applying data science to solve governmental problems and the use of multi-modal data in understanding social phenomenon. His e-mail address is \email{icruicks@andrew.cmu.edu}.\\

\noindent {\bf \MakeUppercase{Jevin West}} is the co-founder of the new Center for an Informed Public at UW aimed at resisting strategic misinformation, promoting an informed society and strengthening democratic discourse. His research and teaching focus on the impact of data and technology on science and society, with a focus on slowing the spread of misinformation. His e-mail address is \email{ jevinw@uw.edu } and his website is \url{https://www.jevinwest.org/}.\\

\end{document}